# Counterions release from electrostatic complexes of polyelectrolytes and proteins of opposite charge : a direct measurement

Jérémie Gummel, Fabrice Cousin* and François Boué

*Laboratoire Léon Brillouin, CEA Saclay 91191 Gif-sur-Yvette Cedex France*

**RECEIVED DATE (automatically inserted by publisher)**; e-mail: fabrice.cousin@cea.fr

There has been recently a large interest [1] for the understanding of the mechanisms governing the complexation of polyelectrolytes and proteins of opposite charges due to the growing potential of applications (proteins fractionation [2], controlled drug release [3], biosensors [4]). This is also linked to the emergence of new polyelectrolyte assemblies such as spherical brushes or multilayers. While an obvious and confirmed driving force of complexation is electrostatic attraction, the role of the counterions (e.g. the Sodium cations of a polyanion chain, noted c.i. below) is a key point. Being many and of small mass, free c.i. have a high translational entropy: e.g., it increases strikingly the osmotic pressure of polyions solutions. If they are dispersed into the solvent after complexation, the entropy gain will be important, and could balance the loss of conformational entropy of the polymer. To estimate this entropy, one has to account for the "condensation" phenomenon: if the distance $a$ between two charges is shorter than the Bjerrum length above which thermal agitation is higher than electrostatic attraction, $l_B$, a large fraction $(1- a / l_B)$ of c.i. is trapped close to the polyion [5] ; condensation is relevant also for spherical objects [6] like globular proteins. During complexation, the charge neutralization of opposite charged species can lead to a release of the condensed c.i. The released c.i regain as much translational entropy as the free ones [7,8]. Such effects could differentiate strongly charged synthetic polyelectrolyte ($a/l_B$ ~2.4Å /7.2 Å ~ 1/3) where condensation is important, from less charged chains (f=1) like polysaccharides. In numerical simulations [9], condensed and free c.i. release is observed. But experiments gave until now only indirect indications: calorimetric measurements during complexation (proteins/polycation [10] or polyanion/polycation [11]) evidence an endothermic entropically driven contribution. Protein penetration within a polyelectrolyte brush [12] is also indirectly attributed to c.i. release. The present paper reports the first direct structural observation, at our best knowledge.

We present here a 'yes - no' experiment, unambiguously linked to the presence or not of the chain c.i. in proteins-polyelectrolyte complexes. The protein is lysozyme, positively charged at low pH, the polyion is polystyrene sulfonate (PSS), with one negative charge per unit. This experiment is based on SANS measurements where we take advantage of the fact that hydrogenated PSS chains and lysozyme have exactly the same neutron density length. Thus the scattering from both species can be switched off simultaneously using a solvent, here a 57%$H_2O$/43%$D_2O$ mixture, which has the same neutron density length: this is called "matching". The remaining signal is the scattering of the counterions only for such "CI" labeling. We chose 4 samples who should, or not, have only released counterions, hence who should scatter, or not.

Our choice of samples is based on already known structures, observed by SANS [13] in complexes of lysozyme and short Sodium PolystyreneSulfonate (PSSNa) which are in dilute regime.

When the ratio of negative to positive effective charges brought by components, $[-]/[+]_{intro}$, is close to 1, the system is made of dense globular primary complexes (radius ~ 10 nm) organized at higher scale in a fractal way. Using this time deuterated PSS chains, we could switch off either the protein scattering (PE labeling) or the d-PSS scattering (Prot labeling). For such globular structure, comparison of data for these two contrasts gives accurately the inner composition of the globules. Typical spectra will be given below in the case of TMA c.i., which gives very similar results. Let us first summarize three points: (i) whatever $[-]/[+]_{intro}$, globules have a neutral dense core, i.e. $[-]/[+]_{inner}$ ~ 1 (note that the negative charge concentration $[-]_{intro}$ comprises all units of the polyion, so core neutrality suggests the release of condensed c.i.). (ii) The core is in addition surrounded by a polymer corona in excess of PSS (hairy core). (iii) This gives four typical structures (Figure 1.a): for $[-]/[+]_{intro}$ < 1, "naked" cores plus free protein, for $[-]/[+]_{intro}$ = 1, naked cores, for $[-]/[+]_{intro}$ > 1, hairy cores, for $[-]/[+]_{intro}$ > 2 hairy cores plus free polyions.

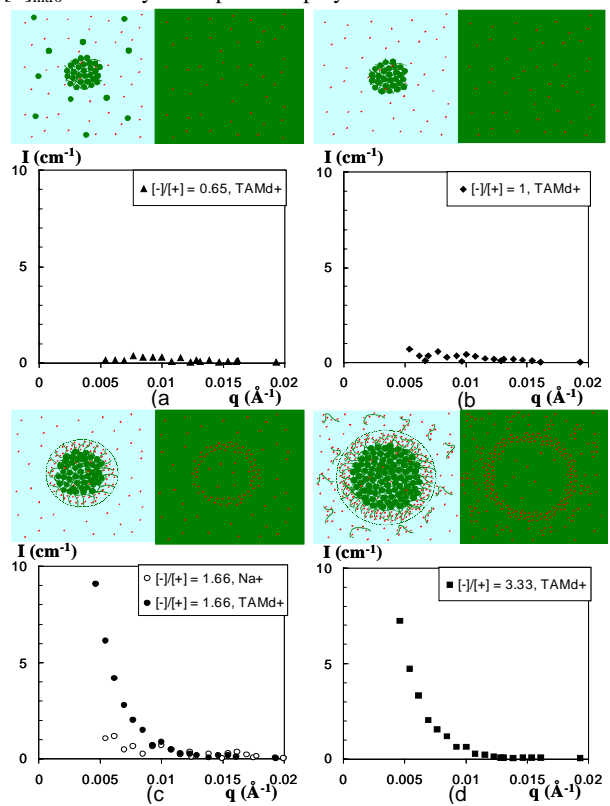

FIGURE 1: Structures of globules made of lysozyme (green), h-PSS (green), and $(CD_3)_4N^+$ counterions (red) in 100%$H_2O$ solvent (left) or in 57%$H_2O$/43%$D_2O$ solvent (right) and corresponding scattering of $(CD_3)_4N^+$ in the 57%$H_2O$/43%$D_2O$ solvent. (a) and (b) : 'naked cores' ((($[-]/[+]_{intro}$ =

0.65 and [-]/[+]$_{intro}$ = 1), (c) and (d): 'hairy cores' ([-]/[+]$_{intro}$ = 1.66 and [-]/[+]$_{intro}$ = 3.33).

Such possibility to tune from "naked" to "hairy" cores permits to compare the scattering where all c.i. should be released - "naked" cores, with the one where some c.i. are still present - "hairy" cores where c.i. are condensed on the dangling polyions of the shell. For a strong conclusion the second should be detectable, while the first should be much lower: using the "CI" labelling conditions, the c.i. scattering would be the one of dispersed very small species, very low [14].

Experimental success relies on the efficiency of the "CI labelling", i.e. on good contrast of counterions: they have to be visible when lysozyme and hydrogenated PSS are masked by the 57%$H_2O$/43%$D_2O$ solvent. The c.i. scattering length density must be very different from the solvent one (in other words have a high contrast). For this we use deuterated tetramethylammonium c.i., $(CD_3)_4N^+$ (noted d-TMA), which contains 12 non labile deuterium atoms (conversely $Na^+$ has -accounting for solvatation, a very poor *contrast* in this solvent [15]).

We firstly checked that replacing Na by TMA does not change the structure of complexes: we performed SANS measurements (PAXY spectrometer, LLB, Saclay, France), with same concentrations as in [13], on samples with d-PSS-TMA. The latter is obtained by Makowski sulfonation [16] of deuterated d-PS chains, like for PSSNa. In the last step, polystyrene sulfonic acid (d-PSSH) is neutralized by tetramethylammoniumhydroxyde (h-TMA(OH)) instead of sodium hydroxide NaOH. The chains weight average polymerization degree N is 40. The complexes are made at pH 4.7 (the protein has a net charge +11) in acetic acid/acetate buffer of ionic strength $5.10^{-2}$ M. Lysozyme is at 40 g/L for all 4 samples, and PSS at 0.02M, 0.03M, 0.05M and 0.1M, corresponding to [-]/[+]$_{intro}$: 0.65, 1, 1.66 and 3.33 respectively. Solvent is either a 57%$H_2O$/43%$D_2O$ mixture that "matches" protein, yielding d-PSS signal, or a 100% $D_2O$ mixture that "matches" d-PSS and yields the protein signal. Results for [-]/[+]$_{intro}$ = 1.66 and 3.33 are shown on figure 2. For a given [-]/[+]$_{intro}$, PSS and lysozyme signals present the same features, indicating that the two species are spatially organized in the same way in the system [13]. We observe: a correlation peak at 0.2 Å$^{-1}$ corresponding to the contact distance between two proteins, a $q^{-4}$ decay at intermediate q due to surface scattering of the globules (which look dense at the corresponding larger scale), and an upturn towards a $q^{-2.1}$ decay at low q due to the fractal organization of the globules at even larger scale. For [-]/[+]$_{intro}$ > 1, the upturn onset is at lower q for the PSS signal because the globule looks larger when the visible species is PSS and a shell is present. Values are lower for d-PSS signal than lysozyme signal because the volume fraction inside the globules when [-]/[+]$_{inner}$ = 1 is lower for PSS than for lysozyme. Thus features are all the same for h-TMA c.i. than with $Na^+$ [13]. The only small difference is the primary complexes size which is slightly larger with TMA counterions.

Then, the specific signal of TMA c.i. was measured using four samples in the 57%$H_2O$/43%$D_2O$ solvent at same concentrations, but replacing d-PSS h-TMA by h-PSS d-TMA. The latter is obtained by neutralization of h-PSSH by d-TMA(OH) (Eurisotop France). On Figure 1, at first sight, samples containing 'naked' globules ([-]/[+]$_{intro}$ = 1 and 0.65) *do not scatter* at low q, while samples with 'hairy' globules ([-]/[+]$_{intro}$ = 1.66 and [-]/[+]$_{intro}$ = 3.33) *do scatter*. A control sample with $Na^+$ c.i. has been measured for [-]/[+]$_{intro}$ = 1.66. It is compared on figure 1 with the d-TMA and does not show any scattering at low q; this confirms that the low q scattering from hairy globules comes from the d-TMA c.i..

In order to check whether this scattering comes from the decoration of the shell by the c.i., we compare it to the PSS chains scattering on figure 2 Applying a renormalization factor F of the order of 10 (11 for [-]/[+]$_{intro}$ = 1.66 and 12 for [-]/[+]$_{intro}$ = 3.33), the c.i. signal fits quite well the PSS signal at low q. One should expect the shape of the curves to be different since corona (for counterions) and sphere (for PSS) have different form factor; however the differences should lie at larger q-range (q > 0.02 Å$^{-1}$), where the c.i. signal is too noisy to be significant. A quick calculation gives an order of the magnitude of the normalization factor F. It should be proportional to the squared ratio of the d-TMA and d-PSS contrasts (~ 1, the neutron density length of $(CD_3)_4N^+$ and d-PSS are close) times the square ratio of their volume fraction ((1-f). $V_{c.i.}$ /$V_{PSS})^2$. Using $1 - f = 1 - (a / l_B) \sim 2/3$), and our estimate of $V_{c.i.}$ /$V_{PSS}$, the calculated F is not very far from the experimental value of about 10. This is a good support of the decoration of the PSS shell by counterions.

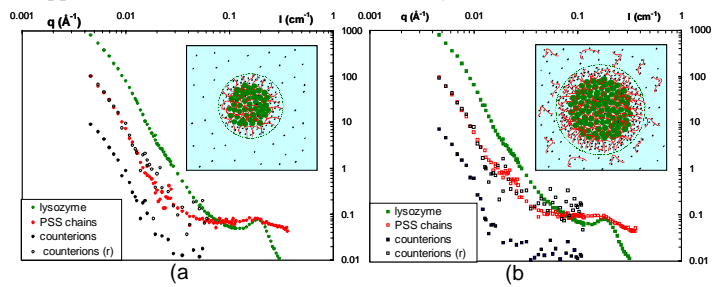

FIGURE 2: Comparison of lysozyme scattering (in 100%$D_2O$ solvent), PSS scattering (d-PSS in 57%$H_2O$/43%$D_2O$ solvent) and c.i. $(CD_3)_4N^+$ in 57%$H_2O$/43%$D_2O$ solvent with h-PSS chains) in hairy globules. The c.i. scattering is also compared with PSS one after normalization. (a) [-]/[+]$_{intro}$ = 1.66, (b) [-]/[+]$_{intro}$ = 3.33

In summary, we have shown by a specific labelling experiment that in a system where proteins and polyelectrolyte chains strongly interact to form dense globules, the inner charge stoichiometry of the globules is accompanied by a complete release of all counterions (condensed as well as not condensed) from the core of the globules (same concentration everywhere in the sample). The only significant scattering signal from c.i. comes from the ones trapped in the shell surrounding the complexes when they exist.

Abstract
Though often considered as one of the main driving process of the complexation of species of opposite charges, the release of counterions has never been experimentally directly measured. We present here the first structural determination of such a release by Small Angle Neutron Scattering in complexes made of lysozyme, a positively charged protein and of PSS, a negatively charged polyelectrolyte. Both components have the same neutron density length, so their scattering can be switched off simultaneously in an appropriate "matching" solvent; this enables determination of the spatial distribution of the single counterions within the complexes. The counterions (including the one subjected to Manning condensation) are expelled from the cores where the species are at electrostatic stoichiometry.

TOC

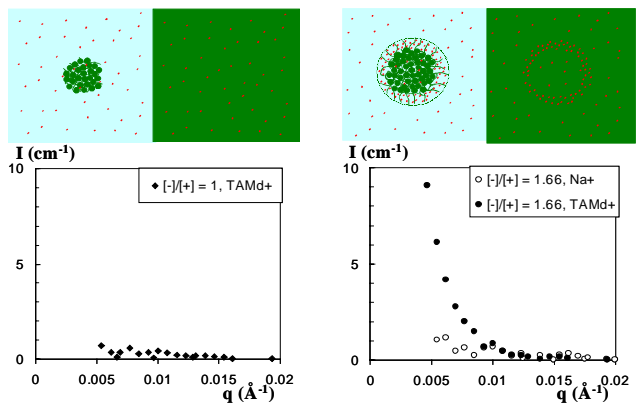